\documentclass[aps,prc,twocolumn,superscriptaddress,showpacs]{revtex4}

\usepackage{graphicx}
\usepackage{latexsym}
\usepackage{bm}
\usepackage{amsmath}

\usepackage{color}

\newcommand{\comment}[1]{}

\begin{document}

\title{Back-to-back correlations of high $p_{\mathrm{T}}$ hadrons
in relativistic heavy ion collisions}

\author{Tetsufumi Hirano}
\affiliation{Physics Department, University of Tokyo,
  Tokyo 113-0033, Japan}
\affiliation{%
RIKEN BNL Research Center, Brookhaven National Laboratory,
                Upton, N.Y. 11973}
\author{Yasushi Nara}
\affiliation{%
RIKEN BNL Research Center, Brookhaven National Laboratory,
                Upton, N.Y. 11973}
\affiliation{%
Department of Physics, University of Arizona, Tucson, Arizona 85721
}

\date{\today}
 
\begin{abstract}
We investigate the suppression factor
and the azimuthal correlation function
for high $p_{\mathrm{T}}$ hadrons
in central Au+Au collisions at $\sqrt{s_{NN}}=200$ GeV
by using a dynamical model in which hydrodynamics
is combined with explicitly traveling jets.
We study the effects of parton energy loss in a hot medium,
intrinsic $k_{\mathrm{T}}$ of partons in a nucleus,
 and $p_{\perp}$ broadening of jets
on the back-to-back correlations of high $p_{\mathrm{T}}$ hadrons.
Parton energy loss is found to be a dominant effect on
the reduction of the away-side peaks in the correlation function.
%it is insufficient to vanish them.
%Intrinsic $k_{\mathrm{T}}$
%and $p_\perp$ broadening
%also reduce the height of the peaks
%by about 10\%  
%within our model.
% On the other hand, the effect of
% initial state intrinsic $k_{\mathrm{T}}$ and
% final state $p_\perp$ broadening
% on the away side peak
% are found to be small.
\end{abstract}

\pacs{24.85.+p,25.75.-q, 24.10.Nz}

\maketitle

%\section{Introduction} 
Recently,
from the two-particle azimuthal correlation measurements
in central Au+Au collisions 
at the Relativistic Heavy Ion Collider (RHIC),
the magnitude of the away-side jets is found to be
significantly suppressed in comparison with the near-side jets
at high $p_{\mathrm{T}}$~\cite{star_btob}.
In addition to the suppression of the yields for high $p_{\mathrm{T}}$
hadrons already discovered at RHIC~\cite{phenix_pi0,star_highpt},
which was predicted some time ago~\cite{Gyulassy:1990ye},
the high $p_{\mathrm{T}}$ correlation measurement may provide
a novel opportunity to study the dense QCD matter
produced in high energy nuclear collisions.
This phenomenon was first predicted by Bjorken in many years ago
\cite{Bjorken:1982tu}. Later, the acoplanarity
(transverse momentum imbalance) of jets
was discussed as a possible
diagnostic tool for studying the quark gluon plasma (QGP) \cite{Appel}.

In this Letter,
 we investigate the effects of
(a) parton energy loss during propagation through an expanding hot medium, 
(b) intrinsic $k_{\text{T}}$ of partons in a nucleus,
and (c) $p_{\perp}$ broadening of jets
on back-to-back correlations in Au+Au collisions
based on incoherent production of mini-jets
by employing the hydro+jet model \cite{HIRANONARA}.

%\section{Model Description} 

The hydro+jet model is mainly composed of two models.
One is a full three dimensional hydrodynamic model
which describes the space-time evolution of 
thermalized partonic/hadronic matter \cite{Hirano:2001eu}.
The other describes the dynamics
(production, propagation, and fragmentation)
of high $p_{\mathrm{T}}$
partons which interact with thermalized partonic matter
through a phenomenological model
for parton energy loss.
A novel hydrodynamic calculation with
the early chemical freeze-out shows
that the $p_{\mathrm{T}}$ slope for pions is insensitive to
thermal freeze-out temperature $T^{\mathrm{th}}$
\cite{HiranoTsuda} and that the $p_{\mathrm{T}}$ spectrum
starts to deviate from the data
at $p_{\mathrm{T}} = 1.5$-$2.0$ GeV/$c$ \cite{HIRANONARA2}.
This result naturally leads us to include
mini-jet components into hydrodynamic results.
Thus we simulate the space-time evolution
of both a fluid and mini-jets \textit{simultaneously}.

%%%%%%%%%%%%%%%%%%%%%%%%%%%%%%%%%%%%%%%%%%%%%%%%%%%%%%%%%%%%%%%%%%%%%%%%%%%%%
% FIG  dN/deta of charged hadrons
%%%%%%%%%%%%%%%%%%%%%%%%%%%%%%%%%%%%%%%%%%%%%%%%%%%%%%%%%%%%%%%%%%%%%%%%%%%%%
% 
% \begin{figure}[t]
% \includegraphics[width=3.3in]{DNDHCGDB20.eps}
% \caption{Pseudorapidity distribution of charged hadrons at
% $\sqrt{s_{NN}} = 200$ GeV collisions. 
% Our hydrodynamic result with $b=2.0$ fm is compared
% with the BRAHMS data for centrality of 5\%.
% }
% \label{fig:dndeta}
% \end{figure}
% 
%%%%%%%%%%%%%%%%%%%%%%%%%%%%%%%%%%%%%%%%%%%%%%%%%%%%%%%%%%%%%%%%%%%%%%%%%%%

Initial parameters in the hydrodynamic model are so chosen as to
reproduce 
$dN_{\mathrm{ch}}/d\eta$ 
in Au+Au collisions
at $\sqrt{s_{NN}} = 200$ GeV observed
by the BRAHMS Collaboration~\cite{Bearden:2001qq}.
The transverse profile of the initial energy density is assumed to
scale with the number of binary collisions \cite{Kolb:2001qz}.
For 5\% central collisions, we choose the impact parameter as
$b=2.0$ fm and the maximum initial energy density at the initial
time $\tau_0=0.6$ fm/$c$ as 39.0 GeV/fm$^3$.
For further details on initialization in our hydrodynamic model,
see Ref.~\cite{HiranoTsuda}.
We note that contribution from hard components
(defined by the particles that have a transverse momentum greater
than 2 GeV/$c$ just after initial collisions)
becomes the order of unity
and can be neglected when we tune initial parameters
in the hydrodynamic model.

We include hard partons using the perturbative QCD (pQCD) parton model.
The momentum distribution of hard partons per hard collision
before traversing hot matter may be written as
\begin{eqnarray}
  E{d \sigma_{\mathrm{jet}}\over d^3 p}
  = K\sum_{a,b} \int g(k_{\mathrm{T,a}})dk_{\mathrm{T,a}}^2
 g(k_{\mathrm{T,b}})dk_{\mathrm{T,b}}^2 \nonumber \\
    \times \int f_a(x_1,Q^2)dx_1f_b(x_2,Q^2)dx_2 E{d\sigma_{ab} \over d^3p},
\end{eqnarray}
where $f$ and $g$ are the collinear and transverse parts of
parton distribution functions. $x$ is the fraction of
the longitudinal momentum and
$k_{\mathrm{T}}$ is the intrinsic transverse momentum
of initial partons inside hadron.
The parton distribution functions $f_a(x,Q^2)$ are taken to be
CTEQ5L~\cite{cteq5} with the scale choice $Q^2= (p_{\mathrm{T}}/2)^2$
for the evaluation of parton distribution.
The minimum momentum transfer $p_{\text{T,min}} = 2.0$ GeV/$c$ is assumed.
The summation runs over all parton species and
relevant leading order QCD processes
are included.
Gaussian intrinsic $k_{\mathrm{T}}$ distribution inside hadron
with the width of $\langle k_{\mathrm{T}}^2\rangle = 1$ GeV$^2/c^2$
 is assigned to the shower initiator
 in the QCD hard $2\to2$ processes.
In the actual calculation,
we use PYTHIA 6.2~\cite{pythia} to simulate each hard scattering
and the initial and final states radiation.
In order to convert hard partons into hadrons,
we use an independent fragmentation
model in PYTHIA after hydrodynamic simulations.
A factor $K$ is used for the higher order corrections.
We have checked that this hadronization model with $K=2.5$
provides good agreement
with the transverse spectra of charged hadrons
 in $p\bar{p}$ collisions~\cite{ua1}
and of neutral pions in $pp$ collisions \cite{Torii}
above $p_{\mathrm{T}} = 1$ GeV/$c$ at $\sqrt{s}=200$ GeV.
The number of hard partons is assumed to scale with the number
of hard scattering which is estimated by using Woods-Saxon nuclear density.
This assumption is consistent with the peripheral 
Au+Au collision at $\sqrt{s_{NN}}=130$ GeV~\cite{phenix_pi0}.
The initial transverse coordinate of a parton is specified by the
number of binary collision distribution for two Woods-Saxon distributions.
We neglect the nuclear shadowing effect
because its effect is small at midrapidity
 at RHIC energies~\cite{Wang:2001gv,jamal}.

%%%%%%%%%%%%%%%%%%%%%%%%%%%%%%%%%%%%%%%%%%%%%%%%%%%%%%%%%%%%%%%%%%%%%%%%%%%%%%
% FIG  R_AA
%%%%%%%%%%%%%%%%%%%%%%%%%%%%%%%%%%%%%%%%%%%%%%%%%%%%%%%%%%%%%%%%%%%%%%%%%%%%%%
%
\begin{figure}[t]
\includegraphics[width=3.3in]{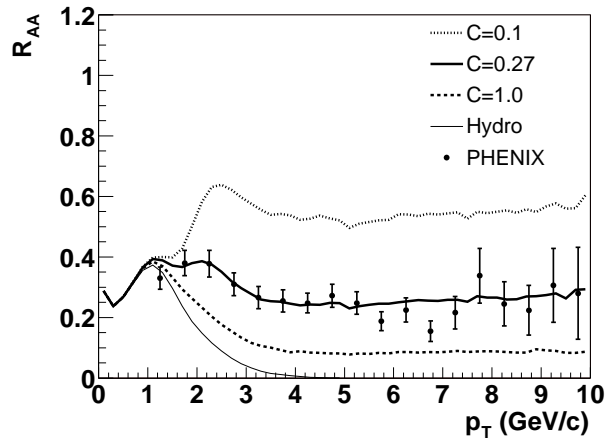}
\caption{Comparison of $R_{AA}(p_{\mathrm{T}})$
for neutral pions
%  defined by Eq.~(\ref{eq:Raa})
in Au+Au 10\% central collision at $\sqrt{s_{NN}} = 200$ GeV.
A parameter $C$ in Eq.~(\ref{eq:GLV}) is chosen as
0.1 (dotted), 0.27 (thick solid), and 1.0 (dashed)
and $\langle k^2_{\mathrm{T}} \rangle=1$ GeV$^2$/$c^2$ is used.
We also show 
%the results from an incoherent model (dash-dotted) and
the hydrodynamic component (thin solid).
Data is taken from Ref.~\cite{Mioduszewski:2002wt}.
}
\label{fig:Raa}
\end{figure}
%
%%%%%%%%%%%%%%%%%%%%%%%%%%%%%%%%%%%%%%%%%%%%%%%%%%%%%%%%%%%%%%%%%%%%%%%%%%%%%%

%%%%%%%%%%%%%%%%%%%%%%%%%%%%%%%%%%%%%%%%%%%%%%%%%%%%%%%%%%%%
% coherent model
%%%%%%%%%%%%%%%%%%%%%%%%%%%%%%%%%%%%%%%%%%%%%%%%%%%%%%%%%%%%

Over the past years, a lot of work
has been devoted to study the propagation
of jets through QCD matter \cite{Baier,Wiedemann,Zakharov,Levai}.
Here we employ the first order formula in opacity expansion
from the reaction operator approach~\cite{Levai,v2jet1}.
The opacity expansion 
is relevant for the realistic heavy ion reactions
where the number of jet scatterings is small.
The energy loss formula for coherent scatterings
in matter
has been applied to analyses of
heavy ion reactions
taking into account the expansion of the system
%where the parton density is dynamically evolved
 \cite{v2jet1,VG}.
The first order formula in this approach is written as
\begin{equation}
\Delta E = C \int_{\tau_0}^{\infty} d\tau
\rho\left(\tau, \bm{x}\left(\tau\right)\right)
(\tau-\tau_0)\ln\left({\frac{2E}{\mu^2 L}}\right).
\label{eq:GLV}
\end{equation}
Here $C$ is an adjustable parameter \cite{note} and
$\rho(\tau,\bm{r})$ is a thermalized parton density.
$\bm{x}\left(\tau\right)$ and $E$ are the position and
the initial energy of a jet, respectively.
In the hydro+jet approach, $\rho$ is obtained
by using hydrodynamic simulations.
We take a typical screening scale $\mu = 0.5$ GeV~\cite{Dumitru:2001xa}
and effective path length $L=3$ fm.
Feedback of the energy to fluid elements
in central collisions was found to be
about 2\% of the total fluid energy.
Hence we can safely neglect its effect on hydrodynamic evolution
in the case of the appropriate amount of energy loss.

%%%%%%%%%%%%%%%%%%%%%%%%%%%%%%%%%%%%%%%%%%%%%%%%%%%%%%%%%%%%
% Results
%%%%%%%%%%%%%%%%%%%%%%%%%%%%%%%%%%%%%%%%%%%%%%%%%%%%%%%%%%%%
%%%%%%%%%%%%%%%%%%%%%%%%%%%%%%%%%%%%%%%%%%%%%%%%%%%%%%%%%%%%%%%%%%%%%%%%%%%%%%
% FIG  Correlation function
%%%%%%%%%%%%%%%%%%%%%%%%%%%%%%%%%%%%%%%%%%%%%%%%%%%%%%%%%%%%%%%%%%%%%%%%%%%%%%
%
\begin{figure}[t]
\includegraphics[width=3.3in]{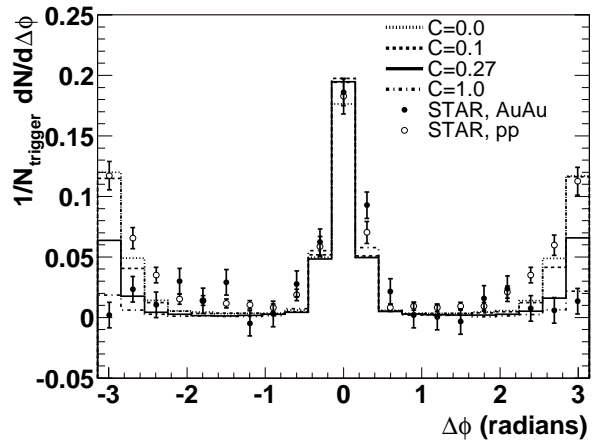}
\caption{Energy loss strength dependence on
$C_2(\Delta \phi)$ for charged particles in central 0-5\% Au+Au collision
 at $\sqrt{s_{NN}} = 200$ GeV.
Parameters $C$ in Eq.~(\ref{eq:GLV}) are 0 (dotted), 0.1 (dashed),
0.27 (solid), and 1.0 (dash-dotted). 
$\langle k^2_{\mathrm{T}}\rangle =1$ GeV$^2$/$c^2$ is used.
Data is taken from Ref.~\cite{star_btob}.
}
\label{fig:c_2}
\end{figure}
%
%%%%%%%%%%%%%%%%%%%%%%%%%%%%%%%%%%%%%%%%%%%%%%%%%%%%%%%%%%%%%%%%%%%%%%%%%%%%%

%First, we discuss the dependence of the magnitude of parton energy loss
%on the suppression factor $R_{AA}(p_{\mathrm{T}})$.
The jet quenching is quantified by the ratio of the particle
yield in $A+A$ collisions to the one in $p+p$ collisions scaled up
by the number of binary collisions
%and the correlation function $C_2(\Delta \phi)$.
%\begin{equation}
$R_{AA} = (d^2N^{A+A}/dp_T d\eta)/(N_{\mathrm{binary}}
d^2N^{p+p}/dp_T d\eta)$.
%\label{eq:Raa}
%\end{equation}
The transverse momentum dependence of $R_{AA}$ can provide information
about the mechanism of jet quenching \cite{jamal,mueller}.
In Fig.~\ref{fig:Raa},
we represent the results of $R_{AA}$ for neutral pions
in Au+Au 10\% central collisions at $\sqrt{s_{NN}}=200$ GeV
from the hydro+jet model
for parameters of $C=0.1$, $0.27$, and $1.0$.
Here we choose $b$ = 2.8 fm for centrality of 10\%.
% For comparison, we also plot a result from
% an model assuming
% $dE/dx \propto \rho$ (which we term an incoherent model).
% We found \cite{HIRANONARA} that
% an incoherent model for parton energy loss
% $dE/dx = 0.06 \rho$ GeV/fm leads us to reproduce the $p_{\mathrm{T}}$
% spectrum of neutral pions at $p_{\mathrm{T}} \sim 3$-$4$ GeV/$c$
% in $\sqrt{s_{NN}} = 130$ GeV central collisions \cite{phenix_pi0}.
% The incoherent model, however, gives the suppression factor which increases
% with $p_{\mathrm{T}}$ above 3 GeV/$c$.
% This behavior is also found in a NLOpQCD study~\cite{jamal}.
% The $p_{\mathrm{T}}$ dependence for the incoherent model is
% completely different from
% the preliminary PHENIX data~\cite{Mioduszewski:2002wt}
% in which $R_{AA}$ gradually decreases with $p_{\mathrm{T}}$.
%On the other hand, the degree of increase
% for the coherent model is rather milder  \cite{VG} than
% for the incoherent model.
The suppression factor for the model with $C=0.27$
quantitatively agrees with the PHENIX data.
%in $p_{\mathrm{T}}<6$ GeV/$c$.
% >From a hydrodynamic simulation, thermalized parton density
% per unit rapidity appears to be $dN_{\mathrm{parton}}/dY = 1274$
% at $\tau_0$.
% Interestingly, a recent prediction of parton density
% from an analysis of $R_{AA}$ \cite{VG}
% is almost consistent with our result.

We now study the dynamical effects of a hot medium
on the strength of away-side peaks in the azimuthal correlation function.
The azimuthal correlation functions
of high $p_{\mathrm{T}}$ charged hadrons
in midrapidity region
($\mid\eta\mid <0.7$) are measured
 by the STAR Collaboration~\cite{star_btob}.
Charged hadrons in
$4 < p_{\mathrm{T, trigger}} < 6$ GeV/$c$ 
and in $2 < p_{\mathrm{T,associate}} < p_{\mathrm{T, trigger}}$
GeV/$c$ are defined to be trigger particles and associated particles
respectively.
The relative azimuthal angle distribution (per trigger particle) 
of high $p_{\mathrm{T}}$ hadrons is written as
\begin{equation}
C_2(\Delta \phi) = \frac{1}{N_{\mathrm{trigger}}}
\int_{-1.4}^{1.4} d\Delta\eta
\frac{dN}{d\Delta\phi d\Delta\eta}.
\end{equation}
Here $\Delta \phi$ and $\Delta \eta$ are, respectively,
the relative azimuthal angle and pseudorapidity between a trigger
particle and an associated particle.
It is found that, in central Au+Au collisions (0-5\%), 
the near-side peak ($\mid \Delta \phi \mid < 0.75$ radians)
has the same strength as the one in $pp$ collisions at $\sqrt{s}=200$ GeV
and the away-side peaks ($\mid \Delta \phi \mid > 2.24$ radians)
disappear clearly~\cite{star_btob}.

In our theoretical calculations, we take into account only contribution from associated particles
which originate from the same hard scattering as
a trigger particle. 
Thus, we compare our results with the STAR data in which 
elliptic flow components are subtracted assuming
$B[1+2v_2^2 \cos(2\Delta\phi)]$ with $B=1.442$ and $v_2=0.07$ \cite{star_btob}.

% We take into account only contribution from jets
% in the correlation analysis
% and simply neglect contribution from hydrodynamic components.
% When we discuss how large a dynamical effect
% on correlation functions is,
% they are normalized in
% background regions
% ($\pi/4 \lsim \mid \Delta \phi \mid \lsim 3\pi/4$).
% Specifically, a constant is added to the correlation function so that
% the sum of each result in the background region coincides
% with that of the STAR data.

%%%%%%%%%%%%%%%%%%%%%%%%%%%%%%%%%%%%%%%%%%%%%%%%%%%%%%%%%%%%%%%%%%%%%%%%%%%%%%
% FIG  intrinsic kT
%%%%%%%%%%%%%%%%%%%%%%%%%%%%%%%%%%%%%%%%%%%%%%%%%%%%%%%%%%%%%%%%%%%%%%%%%%%%%%
\begin{figure}
\includegraphics[width=3.3in]{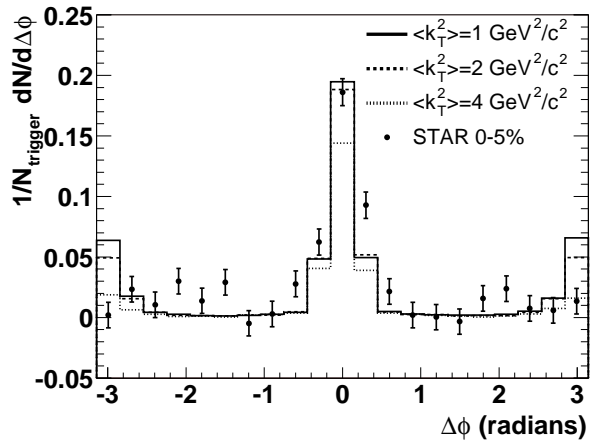}
\caption{Intrinsic $k_{\mathrm{T}}$ dependence of
azimuthal correlation function $C_2(\Delta \phi)$.
Parameter sets are
$(\langle k_{\mathrm{T}}^2 \rangle, C) =(1.0,0.27)$ (solid),
$(2.0,0.35)$ (dashed), and $(4.0,0.5)$ (dotted).
Here the intrinsic transverse momenta
 $\langle k_{\mathrm{T}}^2 \rangle$ are in the unit of GeV$^2/c^2$
and the corresponding energy loss parameters $C$ are chosen
to fit $R_{AA}$.
Data is taken from Ref.~\cite{star_btob}.
}
\label{fig:intrinsickt}
\end{figure}
%%%%%%%%%%%%%%%%%%%%%%%%%%%%%%%%%%%%%%%%%%%%%%%%%%%%%%%%%%%%%%%%%%%%%%%%%%%%%%

As demonstrated in Fig.~\ref{fig:c_2},
% without jet quenching, the strength of
% away-side peaks are the same as the one in $pp$ collisions.
the away-side peaks are gradually decreasing with increasing the amount
of parton energy loss.
Nevertheless, we cannot obtain the disappearance of the away-side peak
by the model with $C=0.27$ which quantitatively reproduces
$R_{AA}(p_{\mathrm{T}})$
%the suppression factor on neutral pions
% at $\sqrt{s_{NN}}=200$ GeV
% in $p_{\mathrm{T}}<6$ GeV/$c$ 
as shown in Fig.~\ref{fig:Raa}.
Only when we underestimate the suppression factor $R_{AA}$ of the PHENIX data
by choosing $C=1.0$ as depicted in Fig.~\ref{fig:Raa},
the away-side peaks almost disappear.
This suggests that another mechanism is needed to
interpret both the suppression factor $R_{AA} \sim 0.25$
and the disappearance
of back-to-back correlation \textit{simultaneously}.
%We will come back to this point in Fig.~\ref{fig:broadening}.

Is a large intrinsic $k_{\mathrm{T}}$ responsible
for the disappearance of back-to-back correlation?
If colliding partons have only longitudinal momenta directed
to the beam axis, produced partons through $2\rightarrow2$ processes
have the same transverse momentum due to kinematics.
When partons in a nucleon initially have the transverse momentum
(intrinsic $k_{\mathrm{T}}$),
this releases the above constraint and, consequently,
affects the back-to-back
correlation of final high $p_{\mathrm{T}}$ hadrons.
Moreover, multiple interactions within the colliding nuclei
lead to the so-called Cronin effect~\cite{Cronin}.
This effect can be parametrized phenomenologically
by increasing the average intrinsic transverse momentum
$\langle k_{\mathrm{T}}^2\rangle$ \cite{XNWang:1998ww,Fai,VG,Dumitru:2001jx}. 
We have chosen in the previous analyses 
$\langle k_{\mathrm{T}}^2\rangle = 1$ GeV$^2/c^2$
which leads to reproduce the $p\bar{p}$ and $pp$ data
at $\sqrt{s}=200$ GeV.
To see the nuclear effect of intrinsic $k_{\mathrm{T}}$ clearly,
we calculate the correlation function by choosing
$\langle k_{\mathrm{T}}^2\rangle = 2$ or $4$ GeV$^2/c^2$.
Here we readjust a parameter $C$ in Eq.~(\ref{eq:GLV}) to quantitatively
reproduce $R_{AA}(p_{\mathrm{T}})$
similar quality to $\langle k_{\mathrm{T}}^2\rangle$ = 1 GeV$^2/c^2$
below 5 GeV/$c$:
$C=0.35$ (0.5) for
$\langle k_{\mathrm{T}}^2\rangle = 2$ (4) GeV$^2/c^2$.  
Only when $\langle k_{\mathrm{T}}^2\rangle = 4$ GeV$^2/c^2$
which is considerably larger
than an estimated value of $\sim 2$-$3$ GeV$^2$/$c^2$
\cite{XNWang:1998ww,Fai,VG,Dumitru:2001jx} at fixed target energies,
the height of the away-side peak becomes small ($\sim 0.02$)
as shown in Fig.~3.
Instead, the near-side peak is also reduced and deviates from data.
%we can obtain the disappearance of away-side peaks.

Measurements in $d+$Au collisions
should provide an important information on
the intrinsic $k_{\mathrm{T}}$ at the RHIC energy region.
The initial $k_{\mathrm{T}}$ broadening should depend on the transverse
position of the hard scattering.
We have checked this effect by using the model in Ref.~\cite{XNWang:1998ww}
which gives average
 $\langle k_{\mathrm{T}}^2 \rangle \sim 2$ GeV$^2/c^2$
for central Au+Au collisions
at RHIC.
We found that the effect of the transverse position dependence
of the intrinsic $k_{\mathrm{T}}$ on the away-side peak is
small and conclusion here remains the same.
It is needed to use the impact parameter dependence of
the intrinsic $k_{\mathrm{T}}$
for study of the centrality dependence of the back-to-back correlation.
One should fix the intrinsic $k_{\mathrm{T}}$ by
$pp$ and $pA(dA)$ data in the systematic study \cite{XNWang:1998ww,Fai}.
% As done by Refs.~\cite{XNWang:1998ww,Fai},
% the intrinsic $k_{\mathrm{T}}$ should be fixed by the experimental data
% of $pp$ and $pA(dA)$.

%%%%%%%%%%%%%%%%%%%%%%%%%%%%%%%%%%%%%%%%%%%%%%%%%%%%%%%%%%%%%%%%%%%%%%%%%%%%%%
% FIG  broading of pt
%%%%%%%%%%%%%%%%%%%%%%%%%%%%%%%%%%%%%%%%%%%%%%%%%%%%%%%%%%%%%%%%%%%%%%%%%%%%%%
\begin{figure}
\includegraphics[width=3.3in]{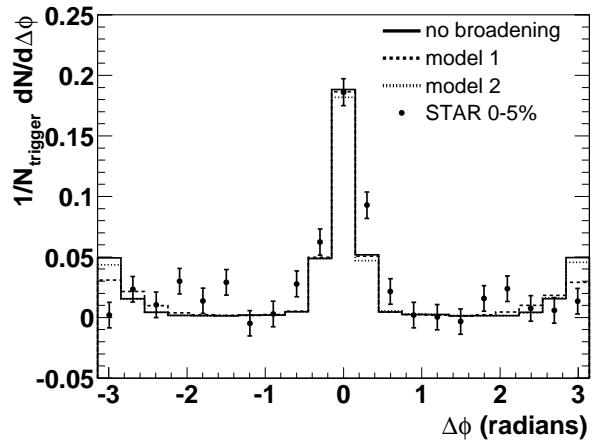}
\caption{
The effect of $p_{\perp}$ broadening on azimuthal
correlation functions $C_2(\Delta \phi)$.
Solid line represents the result for $C=0.35$ and
$\langle k_{\mathrm{T}}^2 \rangle = 2.0$ GeV$^2/c^2$
without $p_\perp$ broadening.
Dashed (dotted) line corresponds to the broadening of model 1 (2).
See text for details.
Data is taken from Ref.~\cite{star_btob}.
}
\label{fig:broadening}
\end{figure}
%%%%%%%%%%%%%%%%%%%%%%%%%%%%%%%%%%%%%%%%%%%%%%%%%%%%%%%%%%%%%%%%%%%%%%%%%%%%%%

Finally, we take into account the $p_{\perp}$ broadening of jets.
Both the non-abelian energy loss and the $p_{\perp}$ broadening of the
propagating jets are manifestation of the strong final state
partonic interactions~\cite{Baier,Wiedemann,Zakharov,Levai,Baier2,
XNWANGQM2002}.
Partons can obtain a transverse momentum orthogonal to its direction
of motion in traversing a dense medium.
As a consequence, traversing partons change
their direction at each time step
and follow zig-zag paths in the hot medium.
The $p_{\perp}$ broadening of these partons is
quantitatively related to the parton energy loss.
We compare two models for the average broadening of partons:
 $\langle p_\perp^2\rangle = (\alpha_{\mathrm{s}}N_{\mathrm{c}}/4)^{-1}
dE/dx$ (model 1) \cite{Baier2}
and 
 $\langle p_\perp^2\rangle 
= (\alpha_{\mathrm{s}}N_{\mathrm{c}}/2)^{-1}
C\int_{\tau_0} d\tau \rho\left(\tau, \bm{x}(\tau)\right)$ (model 2)
\cite{XNWANGQM2002}
where $\alpha_{\mathrm{s}}=0.3$ is assumed.
% According to Ref.~\cite{XNWANGQM2002}, the average broadening of partons
% is proportional to the inverse mean free path and becomes
% $\langle p_\perp^2\rangle 
% = (\alpha_{\mathrm{s}}N_{\mathrm{c}}/2)^{-1}
% C\int_{\tau_0} d\tau \rho\left(\tau, \bm{x}(\tau)\right)$,
% where $\alpha_{\mathrm{s}}=0.3$ and $N_{\mathrm{c}}=3$.
Although the formula  for model 1 is obtained for static plasma,
it is accurate except the factor of logarithm in the case
of 1+1D Bjorken expansion.
Here the distribution of $p_{\perp}$ is assumed
to be a Gaussian function.
It should be noted that incorporation of $p_{\perp}$ broadening
does not largely affect the suppression factor $R_{AA}$.
In Fig.~\ref{fig:broadening},
 the correlation functions  $C_2(\Delta \phi)$
with and without $p_{\perp}$ broadening
are shown for $C=0.35$ and $\langle k_{\mathrm{T}}^2 \rangle=2$ GeV$^2/c^2$.
When the model 1 is used,
 the event average broadening is found to be
$\langle\langle p^2_{\perp}\rangle\rangle \sim 2.5$ GeV$^2/c^2$
 and away-side peaks are suppressed and broadened.
On the other hand, $\langle\langle p^2_{\perp}\rangle\rangle
\sim 0.78$ GeV$^2/c^2$ for model 2 and the reduction
of away-side peaks becomes tiny.
This is consistent with the recent study based on
a pQCD parton model~\cite{Wang:2003mm}.
% There are still small away-side peaks
% for $\langle k_{\mathrm{T}}^2 \rangle = 1$ and 2 GeV$^2$/$c^2$
% if only the longitudinal energy loss of partons
% is taken into account, i.e., partons move along straight paths.
% However, when jets obtain the transverse
% ``kicks" which are perpendicular to its momentum direction, 
% the away-side peaks are weakened for
% $\langle k_{\mathrm{T}}^2 \rangle = 1$ GeV$^2$/$c^2$
% and disappear for $\langle k_{\mathrm{T}}^2 \rangle =$ 2 GeV$^2$/$c^2$.
% For the parameter set of $C=0.3$ and
% $\langle k_{\mathrm{T}}^2 \rangle =$ 2 GeV$^2$/$c^2$,
% the reduction rate of the relative height of away-side peaks
% becomes as much as $\sim 60\%$ from the energy loss and
% the intrinsic $k_{\mathrm{T}}$,
% and $\sim 40\%$ from the $p_{\perp}$ broadening.
% Therefore all three effects (energy loss, 
% nuclear intrinsic $k_{\mathrm{T}}$, and
% $p_{\perp}$ broadening) appear to be the keys for the disappearance
% of away-side peaks in the azimuthal correlation functions.
%\Red{We should comment on the X.N.Wang's paper~\cite{Wang:2003mm}.}

%\section{Summary}

In summary, we have computed the correlation function $C_2(\Delta\phi)$
for high $p_{\mathrm{T}}$ charged hadrons
in central Au+Au collisions at $\sqrt{s_{NN}}=200$ GeV
 within the hydro+jet model.
We have systematically studied the dependence of the magnitude of parton energy
loss on the height of away-side peaks in correlation functions.
Moreover, we also investigated
the effects of the intrinsic transverse momentum
of initial partons and the $p_{\perp}$ broadening of traversing jets
on back-to-back correlations of high $p_{\mathrm{T}}$ hadrons.
About 50\% of the reduction of away-side peaks
comes from the parton energy loss, but its effect is insufficient
to vanish the peaks completely.
This indicates that the suppression of the away-side peaks
occurs by complex mechanisms.
The intrinsic $k_{\mathrm{T}}$ of partons
in a nucleus also affects the back-to-back correlation.
This is, however, due to the increase of
energy loss parameter $C$ in order to
reproduce $R_{AA}$ rather than the increase of $k_{\mathrm{T}}$
itself.
The $p_{\perp}$ broadening causes 5-20\% reduction of the original height
 of away-side peaks.

We would like to thank 
T.~Hatsuda for fruitful discussions and
I.~Vitev for valuable comments on the formula of parton energy loss.
Special thanks go to A. Dumitru for motivating discussions
and continuous encouragement to this work.
We also thank D.~Hardtke, D.~d'Enterria and H.~Torii
for providing us the experimental data.
The work of T.H. is supported by RIKEN. 
Y.~N.~'s research is supported by the DOE under Contract No.
DE-FG03-93ER40792.
%\vspace*{1pc}


\begin{thebibliography}{99}
\bibitem{note}
The proportional constant $C$ in the original formula \cite{Levai}
depends on 
parton species of both jets and matter.
Here we simply assume $C$ is an averaged value.
\end{thebibliography}
\end{document}